\documentclass{article}
\usepackage{amsmath}    
\usepackage{graphicx,dcolumn,amsmath,amssymb,amsfonts,epsfig,float,subfigure}
\usepackage{amsthm}
\newcommand{\al}{\alpha}

\newcommand{\pa}{\partial}

\newcommand{\om}{\omega}

\numberwithin{equation}{section}

\begin{document}
\begin{titlepage}

\begin{center}

{\Large Periodic orbits for an infinite family of classical superintegrable systems }

\vskip 0.7cm

Fr\'ed\'erick Tremblay\footnote{tremblaf@crm.umontreal.ca},\\
Centre de recherches math\'ematiques and D\'epartement de math\'ematiques et de
statistique,\\
Universit\'e de Montreal,  C.P. 6128,
succ. Centre-ville, Montr\'eal (QC) H3C 3J7, Canada, \\[10pt]

Alexander V. Turbiner\footnote{turbiner@nucleares.unam.mx},\\
Instituto de Ciencias Nucleares, UNAM, A.P. 70-543, 04510 M\'exico,\\[10pt]

 and\\[7pt]

Pavel Winternitz\footnote{wintern@crm.umontreal.ca}, \\
Centre de recherches math\'ematiques and D\'epartement de math\'ematiques et de
statistique,\\
Universit\'e de Montreal,  C.P. 6128,
succ. Centre-ville, Montr\'eal (QC) H3C 3J7, Canada
\end{center}

\vskip .9cm

\begin{abstract}
We show that all bounded trajectories in the two dimensional classical system with the potential $V(r,\varphi)=\omega^2 r^2+ \frac{\al k^2}{r^2 \cos^2 {k \varphi}}+ \frac{\beta k^2}{r^2 \sin^2 {k \varphi}}$ are closed for all integer and rational values of $k$. The period is $T=\frac{\pi}{2\omega}$ and does not depend on $k$. This agrees with our earlier conjecture suggesting that the quantum version of this system is superintegrable.
\end{abstract}

\end{titlepage}

\section{Introduction}

A recent article was devoted to an infinite family of quantum systems on a plane described by the Hamiltonian:
\begin{equation}
\label{H}
 H_k \ =\ -\pa_r^2 - \frac{1}{r}\pa_r -
\frac{1}{r^2}\pa_{\varphi}^2  + \om^2 r^2 + \frac{\al k^2}{r^2 \cos^2 {k \varphi}}
+ \frac{\beta k^2}{r^2 \sin^2 {k \varphi}}\ ,
\end{equation}
where $k$ is an arbitrary real number and $(r,\varphi)$ are polar coordinates \cite{TTW2:2009}. These systems were shown to be exactly solvable and integrable for all values of $k$. Integrability is assured by the existence of a second order integral of motion
\begin{equation}
\label{X_k}
   X_k \ =\ L_3^2 + \frac{\al k^2}{\cos^2 {k \varphi}} +
\frac{\beta k^2}{\sin^2 {k \varphi}}, \quad L_3=-i\partial_{\varphi}
\end{equation}
where $L_3 $ is the two-dimensional angular momentum.

The existence of the integral $X_k$ implies the separation of variables in polar coordinates in the Schrodinger equation in quantum mechanics and also in the Hamilton-Jacobi equation in classical mechanics.

Superintegrability in a two-dimensional system means that one further integral of motion exists. Let us call it $Y_k$.

It was conjectured in our previous article \cite{TTW2:2009} that the system (\ref{H}) is superintegrable for all integer values of $k$ and that the additional integral of motion is of order $2k$. This conjecture was proven for $k=1,2,3$ and $4$. If $\alpha=0$ or $\beta=0$, it was proven for $k=1$ to 6 and 8.  For $k=1,2$ and $3$, the system (\ref{H}) reduces to known superintegrable systems. For $k=1$, this is one of the 4 systems found in \cite{Fris:1965,Wint:1966}, for $k=2$ it was found in \cite{Olshanetsky:1983} and for $k=3$ in \cite{Wolfes:1974}.

In all cases ($k=1,2,3$ and $4$) the quantum Hamiltonian and the other quantum integrals of motion lie in the enveloping algebra of an underlying hidden Lie algebra \cite{Turbiner:1998} and this is presumably true for all integer values of $k$ \cite{TTW2:2009}.

The purpose of this article is to analyze the corresponding classical  system with Hamiltonian:
\begin{equation}
\label{HC}
 H_k \ =p_r^2+\frac{p_{\varphi}^2}{r^2}  + \om^2 r^2 + \frac{\al k^2}{r^2 \cos^2 {k \varphi}}
+ \frac{\beta k^2}{r^2 \sin^2 {k \varphi}}\ ,
\end{equation}
We show that all bounded trajectories for this system are closed and that the motion is periodic for all integer and rational  values of $k$. 

We recall that in general, in  an $n$-dimensional space, maximal superintegrability means that the classical Hamiltonian allows $2n-1$ functionally independant integrals of motion (including the Hamiltonian) that are well defined functions on phase space. In classical mechanics maximal superintegrability implies that all bounded trajectories are closed and the motion is periodic \cite{nekhoro}. The corresponding statement in quantum mechanics is that the existence of a non-abelian algebra of integrals of motion implies that the energy levels of the system are degenerate. In the two dimensional case that means that the energy depends on one quantum number, the wave function on two. This is indeed the case for the Hamiltonian (\ref{H}) for all integer values of $k$ and is part of the justification of our conjecture \cite{TTW2:2009}.

In Section 2 we integrate the Hamilton-Jacobi equation for arbitrary values of $k$ in polar coordinates, restricting to the case of bounded trajectories. In Section 3 we obtain an explicit expression for the bounded trajectories in terms of Chebyshev polynomials and show that they are all closed for integer and rational values of $k$. The period in time is $T=\frac{\pi}{2\omega}$. For integer $k$ with $k\geq2$, up to $k-1$ singular points can occur on the trajectories in with the tangent vector is not defined. The special case $k=1$ is discussed in detail in Section 4. This is the only case in which the Hamilton-Jacobi equation separates in a second coordinate system, namely in cartesian coordinates. Section 5 is devoted to examples of trajectories illustrating all features presented in the earlier sections.

\section{Integration of the Hamilton-Jacobi equation}

\subsection{The basic equations}

The Hamilton-Jacobi equation for the considered system in polar coordinates has the form
\begin{equation}
\label{HJ}
H_k= \bigg(\frac{\partial S}{\partial r}\bigg)^2+\frac{1}{r^2}\bigg(\frac{\partial S}{\partial \varphi}\bigg)^2+\omega^2r^2+\frac{1}{r^2}\bigg(\frac{\alpha k^2}{\cos^2k\varphi}+\frac{\beta k^2}{\sin^2k\varphi}\bigg)=E 
\end{equation}
We set the integral of motion (\ref{X_k}) equal to a constant $A$. Since coordinates in (\ref{HJ}) separate we look for a solution in which the classical action has the form
\begin{equation}
 S(r,\varphi,t)=S_1(r)+S_2(\varphi)-Et
\end{equation}
and obtain
\begin{gather}
 r^2\bigg(\frac{\partial S_1}{\partial r}\bigg)^2+\omega^2r^4-Er^2=-A\\
X_k=\bigg(\frac{\partial S_2}{\partial \varphi}\bigg)^2+\frac{\alpha k^2}{\cos^2k\varphi}+\frac{\beta k^2}{\sin^2k\varphi}=A
\end{gather}

Since we are interested in the trajectories we do not need to solve for $S_1$ and $S_2$, though that is obviously possible. Instead we follow the usual procedure \cite{gold} and calculate the derivatives of $S$ with respect to $E$ and $A$. We have
\begin{gather}
\label{S1}
S_1=\int \frac{\sqrt{-A+Er^2-\omega^2r^4}}{r}\ dr\\
\label{S2}
S_2=\int \frac{\sqrt{A\cos^2k\varphi \ \sin^2k\varphi-\alpha k^2\sin^2k\varphi-\beta k^2\cos^2k\varphi}}{\cos k\varphi \ \sin k\varphi}\ d\varphi
\end{gather}
and hence
\begin{align}
\label{delta1} 
\frac{\partial S}{\partial E}&=\frac{\partial S_1}{\partial E}-t=\delta_1\\
\label{delta2} 
\frac{\partial S}{\partial A}&=\frac{\partial S_1}{\partial A}+\frac{\partial S_2}{\partial A}=\delta_2
\end{align}
where $\delta_1$ and $\delta_2$ are constants.

\subsection{Integration of the radial part}

From (\ref{S1}) and (\ref{delta1}) we have
\begin{equation}
 \label{radial}
t+\delta_1=\frac{1}{2}\int\frac{r \ dr}{\sqrt{-A+Er^2-\omega^2r^4}}
\end{equation}
The roots of the denominator are 
\begin{equation}
 (r_{1,2})^2=\frac{1}{2\omega^2}(E\pm\sqrt{E^2-4\omega^2A})
\end{equation}
Bounded motion $r_1\leq r\leq r_2$ is obtained for
\begin{equation}
\label{bounded1} 
E^2-4\omega^2A\geq0\quad\text{and}\quad A\geq 0
\end{equation}

Performing the integration in (\ref{radial}) we obtain
 \begin{gather}
\label{rtime}
r^2=\frac{1}{2\omega^2}\Big(E+\sqrt{E^2-4\omega^2A} \ \sin\big[4\omega(t+\delta_1)\big]\Big)
\end{gather}
Thus for $\omega> 0$, $r^2$ is indeed bounded and periodic with period $T =
\frac{\pi}{2\omega}$ for all values of $\alpha$ and $\beta$ as long as (\ref{bounded1}) is satisfied.

For $\omega =0$ the integral (2.9) must be evaluated differently and we obtain
\begin{equation}
      r^2 = \frac{A+4E^2(t+\delta_1)^2}{E}
\end{equation}
We see that the trajectories for $\omega = 0$ are not bounded, independently
of the values of alpha and beta.

     The limiting case $A=0$ corresponds to the harmonic oscillator for which all
trajectories are bounded and periodic. In the other limiting case $E^2-4 \omega^2A=0$ the integral (2.9) is not defined.
Directly from (2.3) we obtain $\displaystyle\frac{\partial S_1}{\partial r}=0$, $r=\displaystyle\frac{1}{\omega}\sqrt{\frac{E}{2}}$  i.e. the trajectories are
circles.

\subsection{Integration of the angular part}
From (\ref{S1}) and (\ref{S2}) we obtain
\begin{gather}
\frac{\partial S_1}{\partial A}=-\frac{1}{2}\int\frac{dr}{r\sqrt{-A+Er^2-\omega^2r^4}}\\
\frac{\partial S_2}{\partial A}=\frac{1}{2}\int\frac{\sin k\varphi\cos k\varphi \ d\varphi}{\sqrt{A\cos^2k\varphi \ \sin^2k\varphi-\alpha k^2\sin^2k\varphi-\beta k^2\cos^2k\varphi}}
\end{gather}

The integrals are readily evaluated to give
\begin{gather}
\label{S1A}
 \frac{\partial S_1}{\partial A}=-\frac{1}{4\sqrt{A}}\arcsin\bigg[\frac{-2A+Er^2}{r^2\sqrt{E^2-4\omega^2A}}\bigg]\\
\label{S2A}
\frac{\partial S_2}{\partial A}=-\frac{1}{4k\sqrt{A}}\arcsin\bigg[\frac{-2A \ \sin^2k\varphi+A-(\alpha-\beta)k^2}{\sqrt{(A-\alpha k^2+\beta k^2)^2-4k^2\beta A}}\bigg]
\end{gather}

In addition to (\ref{bounded1}) we require
\begin{equation}
 (A-\alpha k^2+\beta k^2)^2-4k^2\beta A>0\quad \text{and}\quad \beta>0
\end{equation}
A more symmetric way of writing this condition is 
\begin{equation}
[A-(\alpha +\beta)k^2]^2-4\alpha\beta k^2\geq0
\end{equation}
From the form of the Hamiltonian (\ref{HC}) we expect the trajectories to be
restricted to sectors
\begin{equation}
 \frac{n\pi}{2k}    <  \varphi< \frac{(n+1)\pi}{2k},\quad n\in\mathbb{R}
 \end{equation}

 The integral (2.15) was evaluated using the substitution $z = \sin^2 k\varphi$. The
denominator in the integral vanishes for the values
\begin{equation}
 z_{1,2}=\frac{(A-\alpha k^2+\beta k^2)\mp\sqrt{(A-\alpha k^2+\beta k^2)^2-4A\beta k^2}}{2A}
\end{equation}

Condition (2.18) is necessary to assure  $z_1 <z< z_2$ which is equivalent
to (2.19). A final condition is $0\leq z_{1,2}\leq1$ which implies 
\begin{equation}
 A+k^2(\beta-\alpha)\geq0\quad \text{and}\quad \alpha>0
\end{equation}

\section{The bounded trajectories}
 \subsection{Equation for trajectories for integer values of $k$ }
So far we have established that bounded trajectories for the Hamiltonian (\ref{HC}) are obtained if the constants of the system satisfy
\begin{equation}
\begin{split}
 A>0,\quad\alpha>0,\quad\beta>0,\quad\omega>0,\quad A> k^2|\beta-\alpha|,\\
 [A-(\alpha +\beta)k^2]^2-4\alpha\beta k^2>0,\quad E^2-4\omega^2A>0
\end{split}
\end{equation}

Now let us obtain an equation for the trajectories. From (2.8), (2.16) and (2.17) we have
\begin{equation}
\label{arcsin} 
k\arcsin R+\arcsin U_k=-4k\sqrt{A}\delta_2
\end{equation}
where we have introduced
\begin{equation}
\label{RandU} 
R=\frac{-2A+Er^2}{r^2\sqrt{E^2-4\omega^2A}}\quad\text{and}\quad U_k=\frac{-2A \ \sin^2k\varphi+A-(\alpha-\beta)k^2}{\sqrt{ [A-(\alpha +\beta)k^2]^2-4\alpha\beta k^2}}
\end{equation}
for $r^2$ and $z=\sin^2k\varphi$ in the intervals
\begin{align}
 \frac{1}{2\omega^2}(E-\sqrt{E^2-4\omega^2A})\leq\ &r^2\leq \frac{1}{2\omega^2}(E+\sqrt{E^2-4\omega^2A})
\\
z_1\leq\ &z\leq z_2
\end{align}
where $z_{1,2}$ are as in (2.21).

We have
\begin{equation}
-1\leq  R\leq 1,\quad -1\leq  U_k\leq 1
\end{equation}

Once $R$ is known as a function of $\varphi$ we obtain $r(\varphi)$ as
\begin{equation*}
 r(\varphi)=\sqrt{\frac{2A}{E-R\sqrt{E^2-4\omega^2A}}}
\end{equation*}

We rewrite (\ref{arcsin}) as
\begin{equation}
\label{arccos} 
k\arccos R=-\arccos U_k+C
\end{equation}
with
\begin{equation}
 C=\frac{(k+1)\pi}{2}+4k\sqrt{A}\delta_2,\quad0\leq C\leq 2\pi
\end{equation}
Equation (\ref{arccos}) can be converted into an algebraic equation for $R$ in term of $\sin^2k\varphi$. We use the well-known formula \cite{grad} for the Chebyshev polynomials
\begin{equation}
 T_k(x)=\cos(k\arccos x)=\frac{(x+i\sqrt{1-x^2})^k+(x-i\sqrt{1-x^2})^k}{2}
\end{equation}
for $k\geq1$ and obtain 
\begin{equation}
\label{TkR}
T_k(R)=\frac{(R+i\sqrt{1-R^2})^k+(R-i\sqrt{1-R^2})^k}{2}=U_k\cos C\pm\sqrt{1-U_k^2}\sin C
\end{equation}

We see that we can restrict the values of $C$ to $0\leq C\leq\pi$ since the interval $\pi\leq C\leq2\pi$ will give the same trajectories.

Let us write (\ref{TkR}) as a polynomial equation with real coefficients for $R$ (and thus for $r^2$). We do this separatly for even and odd values of $k$ and obtain
\begin{align}
\label{evenR}
\begin{split} 
&k=2m:\\ 
&T_{2m}=\sum_{l=0}^{m}\binom{2m}{2l}\sum_{j=0}^l(-1)^j\binom{l}{j}R^{2(m-j)}=U_{2m}\cos C\pm\sqrt{1-U_{2m}^2}\sin C\\
 &k=2m+1:\\
&T_{2m+1}=\\
 &\sum_{l=0}^{m}\binom{2m+1}{2l}\sum_{j=0}^l(-1)^j\binom{l}{j}R^{2(m-j)+1}=U_{2m+1}\cos C\pm\sqrt{1-U_{2m+1}^2}\sin C
\end{split} 
\end{align}

The crucial point is that  (3.11) does not contain any multivalued functions. It follows that $R(\varphi)$ and ultimately $r(\varphi)$ is a periodic function with the same period $\tau$ as $\sin^2k\varphi$ namely $\tau=\displaystyle\frac{\pi}{k}$ and
\begin{equation}
 r(\varphi)= r(\varphi+\frac{\pi}{k})
\end{equation}
Hence all bounded trajectories are closed and the motion is periodic.

We stress that $\tau=\displaystyle\frac{\pi}{k}$ is the period  of $r$ as a function of $\varphi$. Both $r$ and $\varphi$ are periodic in time $t$ with the same period, namely $T=\displaystyle\frac{\pi}{2\omega}$ (see (2.12)) and this period does not depend on $k$ and coincides with that of the harmonic oscillator.

Alternative (and equivalent) equations for the trajectories can be obtained. If we solve (3.10) for $U_k$ we obtain 
\begin{equation}
\label{Uk}
 U_k(\varphi)=T_k(R)\cos C\pm\sqrt{1-T_k(R)^2}\sin C
\end{equation}

Since we have $|U_k|\leq1$ we can define $\eta_k$ by putting
\begin{equation}
 \cos(\eta_k\pm C)=U_k
\end{equation}
and rewrite (3.10) as 
\begin{equation}
T_k(R)= \cos\eta_k
\end{equation}

In section 4 we shall plot the bounded trajectories for low values of $k$. The equation for the trajectories then simplify and we have
\begin{align}
 k=1:&\quad R=U_1\cos C\pm\sqrt{1-U_1^2}\sin C\\
 k=2:&\quad2R^2-1=U_2\cos C\pm\sqrt{1-U_2^2}\sin C\\
 k=3:&\quad4R^3-3 R=U_3\cos C\pm\sqrt{1-U_3^2}\sin C\\
 k=4:&\quad8R^4-8R^2+1=U_4\cos C\pm\sqrt{1-U_4^2}\sin C
\end{align}

The trajectories are always closed curves within the sectors (2.20). They can never actually reach the boundaries of the sectors, unless we have $\alpha=0$ or $\beta=0$ (or both). The trajectories degenerate into line segments for $U_k^2=1$ (see (3.10)). The length of the segment is determined by the limits (3.4) and (3.5) which must be imposed.

\subsection{Singular points on the trajectories for $k$ integer}

Let us consider a curve given in polar coordinates. Singular points on the curve are those where the derivative $\displaystyle\frac{dr}{d \varphi}$ is not defined. They satisfy the overdetermined system of equations:
\begin{equation}
 F(r,\varphi)=0,\quad\frac{\partial F}{\partial r}=0,\quad\frac{\partial F}{\partial \varphi}=0
\end{equation}

In our case the equation of the curve is
\begin{equation}
 F_k(r,\varphi)=T_k(R)-U_k\cos C\mp\sqrt{1-U_k^2}=0
\end{equation}
and the derivatives satisfy
\begin{align}
\begin{split}
 \frac{\partial F_{2m}}{\partial r}&=\frac{\partial T_{2m}}{\partial R}\frac{d R}{d r}\\
&=2\sum_{l=0}^{m}\binom{2m}{2l}\sum_{j=0}^l(-1)^j\binom{l}{j}(m-j)R^{2(m-j)-1}\frac{d R}{d r}\\
\frac{\partial F_{2m+1}}{\partial r}&=\frac{\partial T_{2m+1}}{\partial R}\frac{d R}{d r}
\\
&=\sum_{l=0}^{m}\binom{2m+1}{2l}\sum_{j=0}^l(-1)^j\binom{l}{j}\big[2(m-j)+1\big]R^{2(m-j)}\frac{d R}{d r}
\end{split}\\
 \frac{\partial F_{k}}{\partial \varphi}&=\frac{\cos C\sqrt{1-U_k^2}\pm U_k\sin C}{\sqrt{1-U_k^2}}\frac{d U_k}{d \varphi}
\end{align}

From equation (3.3) we see that $\displaystyle\frac{d R}{d r}$ vanishes only for $r\to\infty$ and $\displaystyle\frac{d U_k}{d \varphi}$ only for $\varphi=\displaystyle\frac{n\pi}{2k}$, $n=0,\pm1,\pm2,...$. These values correspond to points that do not lie on a trajectory.

The singular points satisfy $U_k=\pm\cos C$ that is
\begin{equation}
 \sin^2k\varphi=\frac{A+k^2(\beta-\alpha)\pm\sqrt{(A+k^2(\beta-\alpha))^2-4A\beta k^2}\cos C}{2A}
\end{equation}
and one of equations (3.23) for $k$ even or odd respectively.

From (3.23) we see that $\displaystyle\frac{\partial F_{k}}{\partial r}$ is a polynomial of order $k-1$ and will hence have $k-1$ zeros, all of them real. These zeros are actually singular points if the corresponding values of $r$ and $\varphi$ lie on the trajectories (see (3.21)). This in turn depends on the constants $A$, $C$ and $E$ i.e. on the initial conditions (for $\omega, \alpha$ and $\beta$ given). The statement thus is: the number $n_0$ of singular points on the trajectories satisfies
\begin{equation}
 0\leq n_0\leq k-1
\end{equation}

For small values of $k$ these formulas give:

for $k=1:$
\begin{equation*}
\frac{\partial F_{1}}{\partial r}=\frac{d R}{d r}\neq0\quad r_1\leq r\leq r_2
\end{equation*}
There are no singular points.\\

For $k=2:$
\begin{equation*}
 \frac{\partial F_{2}}{\partial r}=4R\frac{d R}{d r}=0\\
\end{equation*}
The possible singular point corresponds to $\displaystyle r=\sqrt{\frac{2A}{E}}$.\\

For $k=3:$
\begin{equation*}
 \frac{\partial F_{3}}{\partial r}=3(4R^2-1)\frac{d R}{d r}=0\\
\end{equation*}
The two possible singular points are $\displaystyle R=\pm\frac{1}{2}$ i.e.
\[
 r^2=\frac{4A}{2E\pm\sqrt{E^2-4\omega^2A}}
\]

For $k=4:$
\begin{equation*}
 \frac{\partial F_{3}}{\partial r}=16R(2R^2-1)\frac{d R}{d r}=0\\
\end{equation*}
The three possible singular points are $\displaystyle R_1=0$ and $R_{2,3}=\displaystyle\pm\frac{1}{\sqrt{2}}$ i.e.
\begin{equation*}
 r=\displaystyle\sqrt{\frac{2A}{E}},\quad r^2=\frac{2\sqrt{2}A}{\sqrt{2}E\mp\sqrt{E^2-4\omega^2A}}
\end{equation*}

\subsection{Equation for trajectories for rational values of $k=m/n$ }
In the last subsection, we obtained the equation for the trajectories for integer values of $k$. We have established that those trajectories for the Hamiltonian (1.3) are bounded if the constants of the system satisfy condtions (3.1). Now let us obtain the trajectories when $k$ is rational.

Here we set $k=m/n$ for $m,n$ integers. The bounded trajectories in this case are obtained under the same conditions on the constants of the system as those specified in the preceding subsection (3.1).

Equation (3.7) becomes
\begin{equation}
 m\arccos R=n\arccos U_k+\tilde{C}
\end{equation}

Using identity (3.9) we obtain
\begin{equation}
 T_m(R)=T_n(U_k)\cos\tilde{C}\pm\sqrt{1-T_n(U_k)^2}\sin\tilde{C}
\end{equation}

Following the same arguments as in subsection 3.1, we see that $R(\varphi)$ and $r(\varphi)$ are periodic functions of the angle $\varphi$ of period $\tau=\displaystyle\frac{\pi}{k}=\frac{n\pi}{m}$. Again all bounded trajectories are closed and the  motion is periodic.

In Section 5 we will present the trajectories for the rational values $k=1/2, 1/3, 3/2$.

\subsection{Comments on the range of the parameters}

The trajectories given by (3.10) or (3.11) (see also (3.16)-(3.19)) and (3.27) depend on three quantities $E$, $A$ and $C$. $E$ and $A$ are values of the energy $H_k$ (2.1) and the integral of motion $X_k$ (2.4). The third integral $Y_k$, responsible for superintegrability has so far not figured explicitly and is related to the constant $C$. The quantities $A$ and $E$ are restricted by relations (3.1) in order for the trajectories to be bounded. We allow $C$ to take values $0\leq C\leq\pi$. The trajectories could also be calculated differently, using all three integrals of motion and just one of the Newton equations of motion. This may impose further conditions on the parameters involved. We discuss this in detail for $k=1$ in section 4.

\section{The special case $k=1$}  

The bounded trajectories for $k=1$ are given by (3.16) with $R(r)$ and $U_1(\varphi)$ defined in (3.3). The case $k=1$ differs from that of all other values of $k$ by the fact that the additional integral of motion  $Y_k$ for $k=1$ is second order in the momenta and hence also leads to the separation of variables in the Hamilton-Jacobi equation. Explicitly we have \cite{Fris:1965,Wint:1966}
\begin{equation}
Y_1=E_1=P_1^2+\omega^2x^2+\frac{\alpha}{x^2}
\end{equation}
so $Y_1$ is the energy related to motion in the $x$ direction (equally well we could have put $Y_1=E_2=E-E_1$ or $Y_1=E_1-E_2$ where $E_2$ is the energy related to motion in $y$ direction). The Hamilton-Jacobi equation in this case is multiseparable. Namely it separates in polar coordinates (for any $k$) and cartesian ones (for $k=1$). For $k=1$ it also separates in elliptic coordinates with an arbitrary focal distance of $0< f<\infty$.

The equations of motion were solved  in cartesian coordinates \cite{Fris:1965,Wint:1966} and the solution corrresponding to bounded trajectories  in the notations of the present article are
\begin{gather}
\label{trajx}
x^2=\frac{E_1}{2\omega^2}+\frac{1}{2\omega^2}\sqrt{E_1^2-4\alpha\omega^2}\sin[4\omega( t+C_1)]\\
\label{trajy}
y^2=\frac{E_2}{2\omega^2}+\frac{1}{2\omega^2}\sqrt{E_2^2-4\beta\omega^2}\sin[4\omega( t+C_2)]
\end{gather}
Above we have $E_1+E_2=E$ and $C_1$, $C_2$ are integration constants. From (\ref{trajx}) and (\ref{trajy}) we see that the trajectories are bounded and closed. They lie inside a rectangle
\begin{align}
\frac{E_1-\sqrt{E_1-4\alpha\omega^2}}{2\omega^2}\leq \ &x^2\leq \frac{E_1+\sqrt{E_1-4\alpha\omega^2}}{2\omega^2}\\
\frac{E_2-\sqrt{E_2-4\beta\omega^2}}{2\omega^2}\leq \ &y^2\leq \frac{E_2+\sqrt{E_2-4\beta\omega^2}}{2\omega^2}
\end{align}

We see that 
\begin{equation}
2\omega\sqrt{\alpha}\leq E_1\leq E-2\omega\sqrt{\beta}
\end{equation}

Eliminating $t$ from (\ref{trajx}) and (\ref{trajy}) we obtain an equation for the trajectories, namely
\begin{align}
\begin{split}
y^2=\frac{E_2}{2\omega^2}+\frac{\sqrt{E_2^2-4\beta\omega^2}}{2\omega^2\sqrt{E_1^2-4\alpha\omega^2}}\Big[&\cos S(-E_1+2\omega^2x^2)\\
&\pm2\omega\sin S\sqrt{-\alpha+E_1x^2-\omega^2x^4}\Big]
\end{split}
\end{align}
for $0\leq S=C_1-C_2\leq\pi$.

This equation is a fourth order polynomial equation involving the coordinates $x$ and $y$. Viewing $x$ as a parameter for the curve we note that it must satisfy (4.5). The curve (4.7) has four branches, each lying entirely within one quadrant of the $(x,y)$ plane. For $\alpha>0$ and $\beta>0$ it can never reach the coordinate axes (see (4.4) and (4.5)).

In the limiting cases when $\sin S=0$ the closed curve (4.7) collapses into a segment of a quadric. For $S=0$ this is a segment of a hyperbola and for $S=\pi$ a segment of an ellipse (always inside the rectangle (4.4) and (4.5)). 

The trajectory (3.16) in polar coordinates can be made more explicit namely
\begin{gather}
 \label{rphi}
 r^2=\frac{2A}{E-\sqrt{E^2-4\omega^2A}\Big(U\cos C\pm\sqrt{1-U^2}\sin C\Big)}
\end{gather}

Again this is a fourth order polynomial curve (in $r$ and $\sin\varphi$) that degenerates into a segment of a quadric for $\sin C=0$. Considering $z=\sin^2\varphi$ as a parameter of the curve we must restrict to the interval (3.4).

The trajectories given by (4.7) and (4.8) must coincide. This gives a relationship between cartesian constants $\{E_1, E, S\}$ and the polar constants $\{A, E, C\}$. We calculate $r(t)$ and $\varphi(t)$ from (4.2) and (4.3) and compare $r(t)$ from (2.12) and $\varphi(t)$ from (3.16). For given total energy $E$ (and $\alpha, \beta, \omega$) we obtain
\begin{equation}
A=\frac{E_1E_2+2(\alpha+\beta)\omega^2-\sqrt{(E_1^2-4\alpha\omega^2)(E_2^2-4\beta\omega^2)}\cos S}{2\omega^2}
\end{equation}
and
\begin{equation}
\cos C=\frac{A(2E_1-E)-E(\alpha-\beta)}{\sqrt{(A^2+(\alpha-\beta)^2-2A(\alpha+\beta))(E^2-4\omega^2A)}}
 \end{equation}

The condition $|\cos C|\leq1$ is not necessarily automatically satisfied by the right hand side of (4.10) and imposes further conditions. First of all, the denominator in (4.10) must be the square root of a positive number which implies
\begin{equation}
A\geq(\sqrt{\alpha}+\sqrt{\beta})^2
\end{equation}

Secondly, the equation $\cos^2C=1$ is a quadratic polynomial in $A$ and $E_1$
\begin{equation}
\omega^2A^2+A(E_1^2-E_1E-2(\alpha+\beta)\omega^2)+E^2\alpha+E_1E(-\alpha+\beta)+(\alpha-\beta)^2\omega^2=0
\end{equation}

An example of the corresponding curve in the $(E_1,A)$ plane is given on Fig. 1. We have choosen $E=20$,  $\omega=2$, $\alpha=3$ and $\beta=2$. The curve $\cos^2C=1$ lies inside the rectangle 
\begin{gather*}
(\sqrt{\alpha}+\sqrt{\beta})^2\leq A\leq \frac{E}{4\omega^2}\\
2\omega\sqrt{\alpha}\leq E_1\leq E-2\alpha\sqrt{\omega}
\end{gather*}

We have $0<\cos C<1$ in region I, $-1<\cos C<0$ in II, $\cos C<-1$ III and IV, $\cos C>1$ in V and VI.
\begin{center}
\begin{figure}[H]
 \includegraphics[width=10cm]{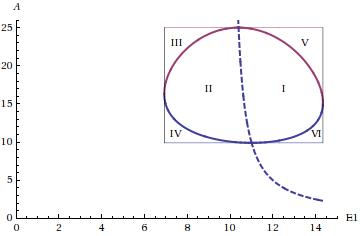}
\caption{Region of bounded trajectories in $(E_1,A)$ plane} 
\end{figure}
\end{center}

The dashed curve represents $\cos C=0$. At the intersection points of this curve and (4.12), we have  $\cos C=\frac{0}{0}$.

\section{Examples of trajectories}

In this section we present curves showing trajectories for specific choices of parameters  $\omega$, $\alpha$ and $\beta$ and the integrals of motion $E$, $A$ and $C$. We choose  $k=1,2,3,4, 1/2, 1/3, 3/2$.

From all of the figures we see that the trajectories do indeed lie within the sectors (2.20) and never  reach the sector boundaries. From formulas in Sections 2 and 3 and from the figures we see that we can actually restrict to one sector only e.g. the one with $n=0$ in (2.20). The trajectories in all others sectors  are obtained by reflections in the sector boundaries.

The figures illustrate all features discussed in the previous sections. For $k=1$, the trajectories have no singular points. The dependance on the constant $C$ is quite spectacular. For all  values of $C$ the trajectories are fourth order curves. For $C=\frac{\pi}{2}$ they resemble ellipses. For $C=0$ or $C=\pi$ the trajectories flatten out and finally degenerate into line segments (see Fig. 2-7). This happens for all values of $k$. For $k=2$, (3.17) has two real roots. For each of them we present the trajectories in all 8 sectors for just one value of $C $ (see Fig. 8 and 9). We see that the chosen trajectories have one singular point.  For $k=3$, (3.18) has three roots but only one is real. A typical set of trajectories is on Fig. 10. For $k=4$, (3.19) has 4 real roots. We choose just one of them and present the trajectories in the first quadrant of the $(x,y)$ plane (see Fig. 11).

Trajectories for rational values of $k$, namely $k=1/2, 1/3, 3/2$ are given on Fig. 12, 13 and 14 respectively. They somewhat resemble Lissajous curves for an anisotropic oscillator. The symmetry with respect to reflection in the boundaries is obvious in Fig. 7-11.

\subsection{k=1}
We set $E=20$, $A=12$, $\omega=2$, $\alpha=3$ and $\beta=2$. The trajectories are given on Fig. 2 to 7.
\begin{figure}[h]
\centering
\begin{tabular}{cc}

\begin{minipage}{2in}
\centering
\includegraphics[width=3.5cm]{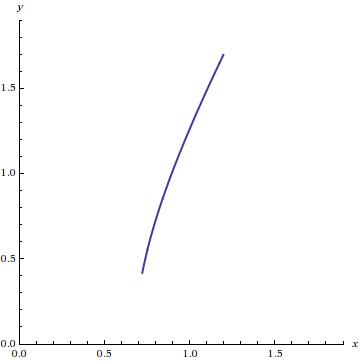}
\caption{$C=\pi$} 
\end{minipage}

&

\begin{minipage}{2in}
\centering
\includegraphics[width=3.5cm]{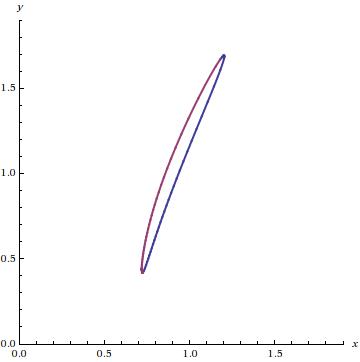}
\caption{$C=\frac{95\pi}{100}$} 
\end{minipage}

\\\\\\
\begin{minipage}{2in}
\centering
\includegraphics[width=3.5cm]{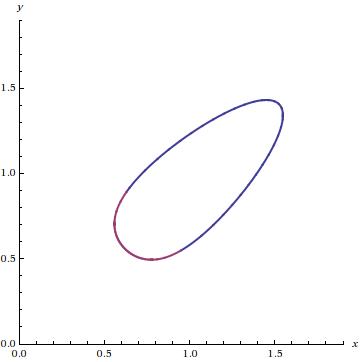}
\caption{$C=\frac{\pi}{2}$} 
\end{minipage}

&

\begin{minipage}{2in}
\centering
\includegraphics[width=3.5cm]{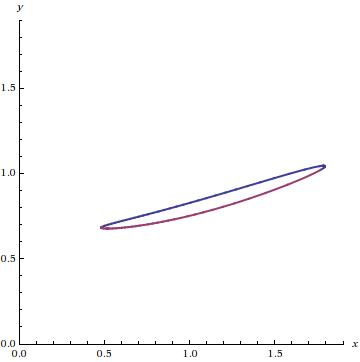}
\caption{$C=\frac{\pi}{20}$} 
\end{minipage}
\end{tabular}
\end{figure}
\newpage
\begin{figure}[h]
\centering
\begin{tabular}{cc}
 \begin{minipage}{2in}
\centering
\includegraphics[width=3.5cm]{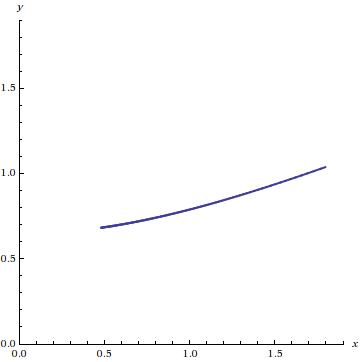}
\caption{$C=0$} 
\end{minipage}
&
 \begin{minipage}{2in}
\centering
\includegraphics[width=3.5cm]{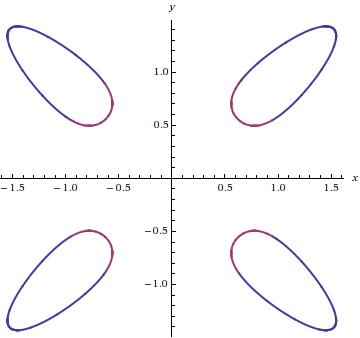}
\caption{Trajectories in the four sectors  of the $(x,y)$  plane for $C=\frac{\pi}{2}$.} 
\end{minipage}
\end{tabular}

\end{figure}

\subsection{k=2}
We set $E=50$, $A=60$, $\omega=2$, $\alpha=3$ and $\beta=2$. The trajectories in the eight sectors  for $C=\frac{\pi}{2}$ are shown on Fig. 8 and 9.
\begin{figure}[h]
\begin{tabular}{cc}
\begin{minipage}{2in}
\centering
\includegraphics[width=5cm]{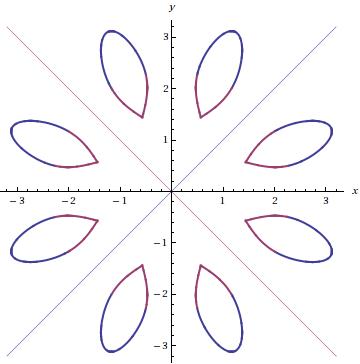}
\caption{Trajectories for the first root}
\end{minipage}
&
\begin{minipage}{2in}
\centering
\includegraphics[width=5cm]{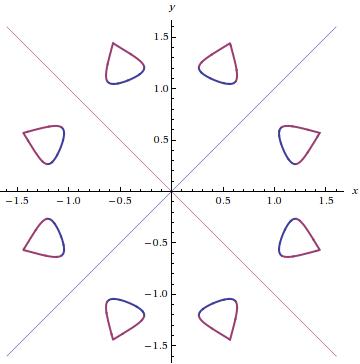}
\caption{Trajectories for the second root}
\end{minipage}
\end{tabular}
\end{figure}
\subsection{$k=3$ and $4$}
 For $k=3$ we set $E=50$, $A=100$, $\omega=2$, $\alpha=3$ and $\beta=2$  and for $k=4$, $E=16$, $A=15$, $\omega=2$, $\alpha=\frac{1}{4}$ and $\beta=\frac{1}{8}$ . The trajectories are  shown on Fig. 10 and 11.
\newpage

\begin{figure}[h]
\begin{tabular}{cc}
\begin{minipage}{2in}
\centering
\includegraphics[width=5cm]{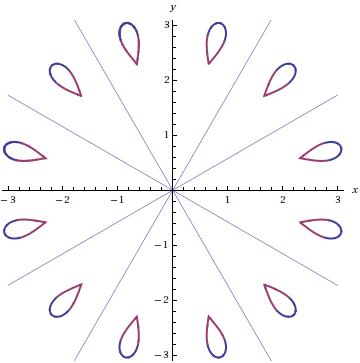}
\caption{Trajectories for the real root for $k=3$}
\end{minipage}
&
\begin{minipage}{2in}
\centering
\includegraphics[width=5cm]{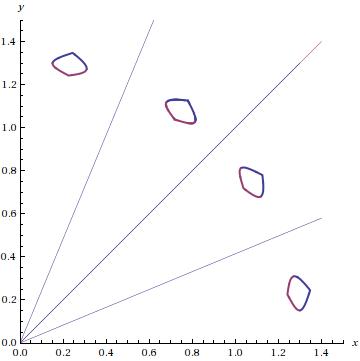}
\caption{Trajectories for the first real root in the first quadrant for $k=4$}
\end{minipage}
\end{tabular}
\end{figure}


\subsection{$k=1/2$ and $1/3$}
In (3.28), we set $E=20$, $A=24$, $\omega=2$, $\alpha=3$ and $\beta=2$.  So for $C=\frac{\pi}{2}$,  the trajectories are given on Fig. 12 and 13. There are no singular points.
\begin{figure}[h]
\begin{tabular}{cc}
\begin{minipage}{2in}
\centering
\includegraphics[width=5cm]{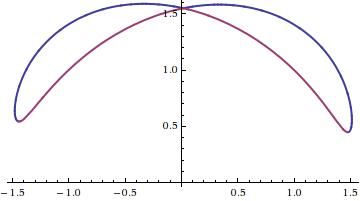}
\caption{Trajectories for $k=\displaystyle\frac{1}{2}$}
\end{minipage}
&
\begin{minipage}{2in}
\centering
\includegraphics[width=5cm]{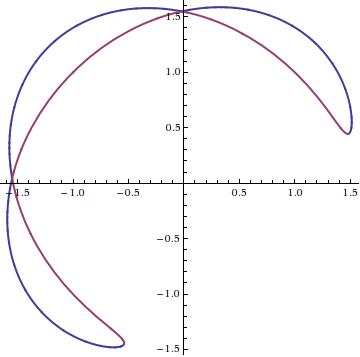}
\caption{Trajectories for $k=\displaystyle\frac{1}{3}$}
\end{minipage}
\end{tabular}
\end{figure}
\newpage
\subsection{$k=\displaystyle\frac{3}{2}$}
From (3.28), we set $E=15$, $A=10$, $\omega=2$, $\alpha=\frac{1}{4}$ and $\beta=\frac{1}{8}$. We obtain three roots for $r^2$. Two of them are complex conjugate to each other. The trajectories are corresponding to the real root are for $C=\frac{\pi}{2}$ are shown on Fig. 14. We observe two singular points.
\begin{figure}[H]
\centering
\includegraphics[width=6cm,height=4cm]{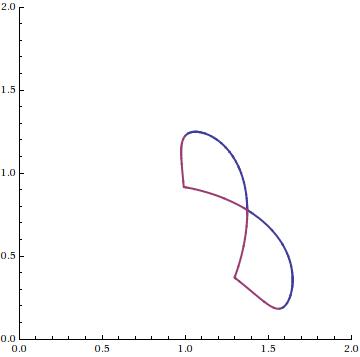}
\caption{Trajectories for $k=\displaystyle\frac{3}{2}$} 
\end{figure}
\section{Conclusion}

The present article, together with our previous one \cite{TTW2:2009}, provides a very strong indication that the Hamiltonian (1.3) is superintegrable for all integer  values of $k$ in classical and in quantum mechanics. In classical mechanics, the results of section 3.3 show that this is also true for rational values of $k$. Indeed we have shown in \cite{TTW2:2009} that the system (1.1) is exactly solvable for all values of $k$ thus supporting a previous conjecture \cite{TTW:2001} that all superintegrable systems are exactly solvable. A spectacular result is that the period $T=\frac{\pi}{2\omega}$ is not only independent of $k$ but is the same as for the pure harmonic oscillator.

We mention that the Hamiltonian (1.3) can be interpreted as representing a three body system on a line, once the motion of the center-of-mass is factored out. The same classical system was studied in \cite{chanu} for $\omega=0$ and $\alpha=0$. In this case the trajectories are never bounded (see (3.1)) and the motion cannot be periodic. A third integral of motion can still exist and the authors  suggest a possible form of the additional integral\cite{chanu}.

A rigorous proof of superintegrability requires the explicit construction of the third integral of motion for all integer and rational   values of $k$ and possibly even for any $k>0$, $k\in\mathbb{R}$. This is left for  a future study.

\section*{Acknowledgments}
The research of A.V.T. is supported in part by DGAPA grant  ININ115709 (Mexico). The research of P.W. was partially supported by a research grant from NSERC of Canada.  We thank Benoit Huard for useful discussions. 
{}


\begin{thebibliography}{}

\bibitem{TTW2:2009} F. Tremblay, A. V. Turbiner and P. Winternitz, ``An infinite family of solvable and integrable quantum systems on a plane'', \emph{J. Phys. A: Math. Theor.\textbf{42}}, 242001 (2009).

\bibitem{Fris:1965}  J.~Fri$\breve{s}$, V.~Mandrosov,
Ya.A.~Smorodinsky, M.~Uhlir and P.~Winternitz, ``On higher symmetries in quantum mechanics'',
\emph{Phys.Lett.\textbf{16}}, 354-356 (1965).

\bibitem{Wint:1966}  P.~Winternitz, Ya.A.~Smorodinsky, M.~Uhlir and
J.~Fri$\breve{s}$,``Symmetry groups in classical and quantum mechanics'', \emph{Yad. Fiz. \textbf{4}}, 625-635
(1966); \emph{Sov. Journ. Nucl. Phys \textbf{4}}, 444-450 (1967)
(English Translation).

\bibitem{Olshanetsky:1983}
    M.~A. Olshanetsky and A.~M. Perelomov, ``Quantum integrable systems related to
Lie algebras'', {\em Phys. Rep.} {\bf 94}, 313 (1983).

\bibitem{Wolfes:1974}
           J.~Wolfes, ``On the three-body linear problem with three-body
interaction," {\em J.~Math. Phys. \bf 15} 1420-1424  (1974).

\bibitem{Turbiner:1998}
           A.~Turbiner, ``Hidden Algebra of Three-Body Integrable Systems'', {\it Mod.Phys.Lett. \bf A13}, 1473-1483 (1998).

\bibitem{nekhoro}
       N.N.Nekhoroshev,
       ``Action-angle variables and their generalization'', {\em Trans. Moscow. Math. Soc. \bf 26}, 180-198 (1972).

\bibitem{gold} H. Goldstein, C. P. Poole and J. L. Safko,``Classical Mechanics'', Addison Wesley  (3rd Edition),  2001.

\bibitem{grad}  I.S. Gradshteyn and I.M. Ryzhik,  ``Table of Integrals, Series and Products'', Academic Press (7th edition), 1980.

\bibitem{TTW:2001}
       P.~Tempesta, A.~Turbiner, P.~Winternitz, ``Exact Solvability of
       Superintegrable Systems", {\it Journ.Math.Phys. \bf 42}, 4248-4257 (2001)

\bibitem{chanu}
       C. Chanu, L. Degiovanni and G. Rastelli,``Superintegrable three-body systems on a line",   {\em J. Math. Phys.}  {\bf 49}, 112901 (2008) (10 pages)


\end{thebibliography}
\end{document}